\documentclass{aa}

\usepackage{graphicx,epsfig,fancyhdr,rotating,amsmath,txfonts,natbib}
\def\kms{\rm km\;s$^{-1}$}

\usepackage[none]{hyphenat}

\begin{document}

\title{Can coronal hole spicules reach coronal temperatures?}

\author{M.S. Madjarska, K. Vanninathan \and J.G. Doyle}
\offprints{madj@arm.ac.uk}
\institute{Armagh Observatory, College Hill, Armagh BT61 9DG, N. Ireland}

 \date{Received date, accepted date}

\abstract
{}
{We aim with the present study to provide observational evidences on whether coronal hole spicules reach coronal temperatures. }
{We combine  multi-instrument  co-observations obtained  with the SUMER/SoHO  and with the  EIS/SOT/XRT/Hinode. }
{The analysed three large spicules were found to be comprised of numerous thin spicules which rise, rotate and descend simultaneously forming a bush-like feature. Their rotation resembles the untwisting of a large flux rope. They show velocities ranging from 50 to 250 \kms. We clearly associated the red- and blue-shifted emissions in transition region lines with rotating but also with rising and descending plasmas, respectively. Our main result is that these spicules although very large and  dynamic, show no  presence in spectral lines formed at temperatures above 300\,000~K. }
{The present paper brings out the analysis of three Ca~{\sc ii}~H large spicules which  are composed of numerous dynamic thin spicules but appear as  macrospicules  in EUV lower resolution images. We found no coronal counterpart of these and  smaller spicules.  We believe that the identification of phenomena which have very different origins as macrospicules is due to the interpretation of the transition region emission, and especially the He~{\sc ii} emission, wherein both chromospheric large spicules and coronal X-ray jets  are present. We suggest that the recent observation of spicules in the coronal AIA/SDO 171~\AA\ and 211~\AA\ channels is probably due to the existence of transition region emission there. }

\authorrunning{Madjarska, M. S. et al.}
\titlerunning{Spicules}

\maketitle

\section{Introduction}
The term spicule refers to  jet-like features expelled from the chromosphere as seen  at the solar 
limb. They were first observed by \cite{Secchi1877} and named spicules by \cite{Roberts1945}. 
Spicules are best viewed at the solar limb as bright features against the dark background of 
the solar corona in H$\alpha$ and Ca~{\sc ii}  images. Several studies report that these 
phenomena fall back along the same trajectory or fade out \citep{Beckers1972, Suematsu1998}. 
Many on-disk filament like features were identified as the counterpart of the limb spicules  
due to the similarities of their properties \citep{Christopoulou2001,Rouppevandervoort2007} and 
some were named `mottles' \citep{Tsiropoula1997}. Coronal hole spicules are found to be taller than quiet Sun spicules, probably 
due to the different configuration of the magnetic field of the two regions \citep{Beckers1972}.
Spicules/mottles have been observed in 
temperatures between 5\,000~K and 300\,000~K. However, \citet{2011Sci...331...55D} reported 
that a small but sufficient fraction of spicules, including coronal hole spicules  are heated to temperatures above 1~MK based on 
observations from the Atmospheric Imaging Assembly (AIA) instrument onboard the Solar 
Dynamic Observatory (SDO) taken with the 171~\AA\ filter. Transient events like spicules/mottles 
are of prime importance as they intermittently connect the chromosphere with the corona and 
possibly sustain the mass balance in the solar atmosphere \citep{Tsiropoula2004}. The 
down-flow observed in transition region lines was suggested to result from the mottle/spicule 
plasma returning to the solar surface  \citep{Tsiropoula2004, Pneuman1978, Withbroe1983}.   
If, however, all the material that is sent up through spicules/mottles is returned to the 
solar surface then their contribution to coronal heating will be minimal, dismissing the 
possibility of spicules directly contributing to coronal heating \citep{Withbroe1983}. 
Moreover, there are speculations that spicules/mottles maybe capable of transporting 
energy high into the upper chromosphere and even up to the corona \citep{Pneuman1978, Athay1982} 
and, therefore, can be considered as possible candidates responsible for coronal heating 
\citep{2000SoPh..197...31A, 2011Sci...331...55D}. 

 \begin{figure*}[ht!]
 \centering
 \vspace{5.5cm}
\includegraphics{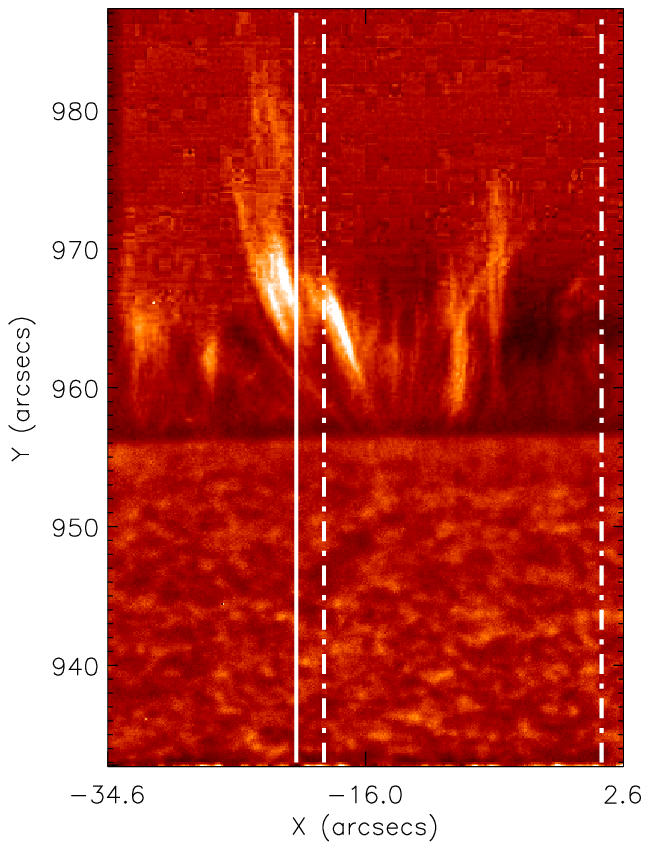}
\includegraphics{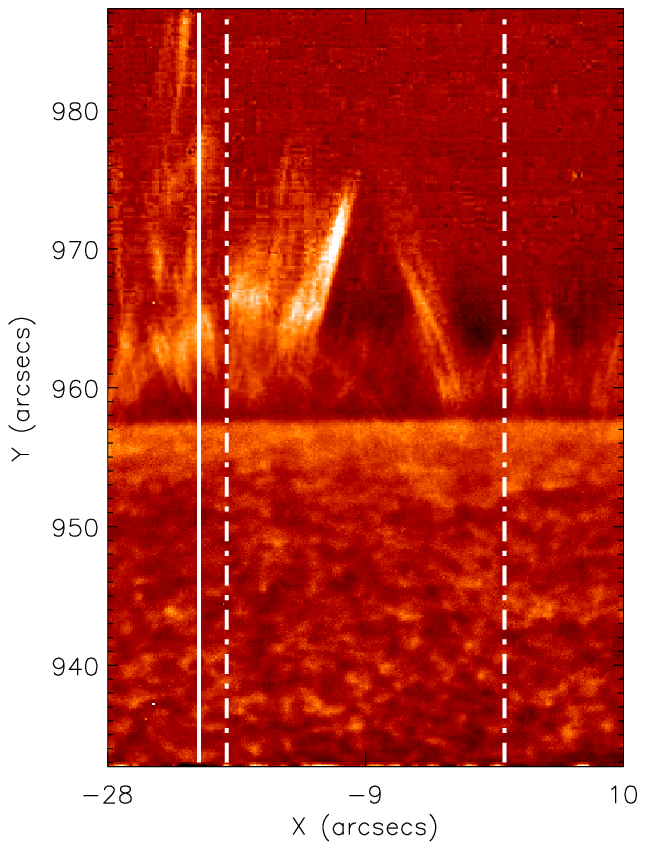}
\includegraphics{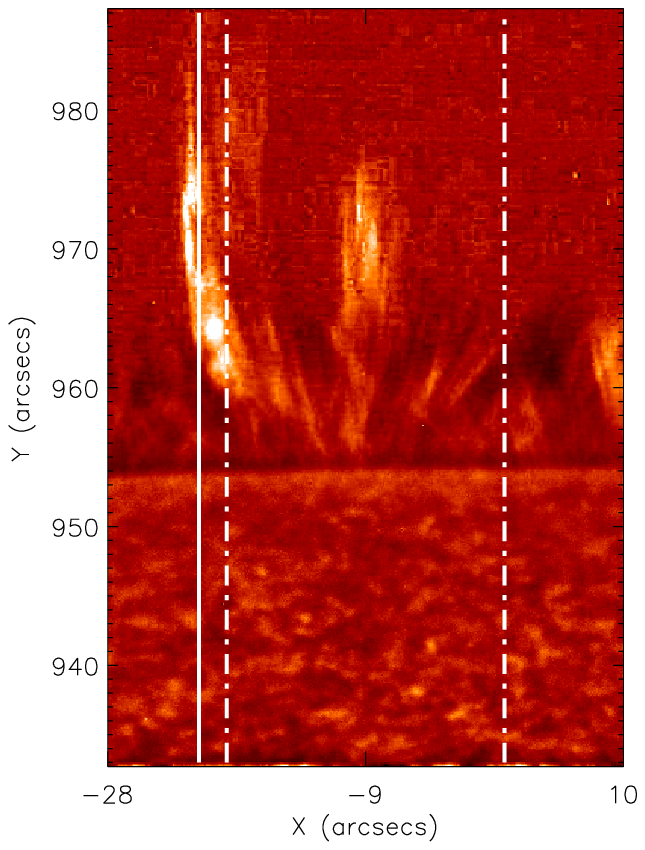}
 \caption{SOT Ca~{\sc ii} images showing spicules taken from left to right, on 2009, April 28 at 16:24~UT, on April 29 at 02:24~UT and 03:28~UT.  The over-plotted solid line indicates the position of the SUMER slit, while the dashed-dotted lines indicate  the EIS raster FOV.}
 \label{fig1}
 \end{figure*}

\citet{1975ApJ...197L.133B} revealed the existence of  jet-like features found in images taken with 
the slitless NRL spectrograph during the Skylab mission. The events were named macrospicules as they 
resembled H$\alpha$ spicules, but were much larger and had longer lifetime. They appear 
increasingly inclined away from the pole as a function of increasing position angle measured from the 
pole which makes them comparable to H$\alpha$ spicules. They were first seen in He~{\sc ii}~304~\AA\ 
and only rarely in Ne~{\sc vii}~465~\AA\  (5.7~K) \citep{1975ApJ...197L.133B}.  We would like to note here 
that the formation temperature of He~{\sc ii} ranges from 5 to 12$\times$10$^4$~K with a maximum at 
8$\times$10$^4$~K, corresponding to a low transition region temperature or the uppermost chromosphere 
\citep{1974A&A....34...69J}.  Due to the anomalous behaviour of the He~{\sc i} and {\sc ii} lines 
\citep{2003A&A...400..737A}, the presence of transient events in these 
lines  does not directly tell where in the solar atmosphere these events originate. Therefore, 
to link features originating in the chromosphere such as spicules with phenomena seen at temperatures 
describing the transition region and corona is not a trivial task and strongly requires the use of suitable data. 
 
Spectral and imaging  macrospicule analysis were reported in several 
papers \citep{1997SoPh..175..457P, 
2002A&A...384..303P, 2002SoPh..206..359P, 2010A&A...510L...1K}.  In SoHO/CDS  data, they were registered in spectral lines with formation temperatures from 20\,000~K to 1~MK \citep{1997SoPh..175..457P}, although \citet{2002SoPh..206..359P} found no Mg~{\sc ix}  (1~MK)  emission in the region where macrospicules were detected in He~{\sc i} (20\,000~K) and O~{\sc v} (250\,000~K).  Recently, \citet{2010A&A...510L...1K} described a macrospicule seen in the He~{\sc ii}~256~\AA\  and  Fe~{\sc xii} 195~\AA\ (1.2~MK) lines from the Extreme-ultraviolet Imaging Spectrometer (EIS) onboard the Hinode satellite. The authors  
associated the macrospicule with an X-ray jet from a coronal bright point but no counterpart was reported in  SOT Ca~{\sc ii}~H. \citet{2010ApJ...722.1644S} presented observations of an X-ray jet in X-ray telescope (XRT) images also seen in EIS  266\arcsec\ slot images consisting of a blend of several lines
including He~{\sc ii} and Fe~{\sc xv}.  The X-ray jet was associated with tall spicules seen in images from the Solar Optical Telescope (SOT) taken with the Ca~{\sc ii}~H filter and identified as type II spicules.

We performed specially planned multi-wavelength  Hinode (EIS/SOT/XRT) and SoHO (Solar Measurements of 
Emitted Radiation (SUMER)) co-observations at the solar limb. {The present paper describes the 
association of simultaneously evolving spicules seen in SOT Ca~{\sc ii} with macrospicules judging from 
their appearance in SUMER and  EIS observations.} The aim 
of our investigation  was to probe the possibility that spicules reach coronal temperatures, i.e. can deposit  thermal energy  directly in the solar corona.  Sect.~2 describes the analysed data and 
their alignment. In Sect.~3 we give the obtained results. In Sect.~4 we discuss the open questions and 
state our conclusions. 

\section{Observations}
The observations were taken  at the North pole on 2009, April 28 and 29. The events were registered 
by the SOT, EIS and XRT onboard Hinode and the SUMER spectrometer onboard SoHO (Fig.~\ref{fig1}). 
The SOT \citep{Tsuneta2008} took observations  with the Ca~{\sc ii}~H filter with a cadence of 
10~s. The EIS \citep{Culhane2007a} observations were done with a 2\arcsec\ slit and a 60~s exposure 
time. They consist of a large raster with a size of 70\arcsec $\times$ 248\arcsec\ followed by 
small rasters of 24\arcsec $\times$ 248\arcsec.  EIS took observations in many spectral lines, 
e.g. Fe~{\sc viii}, Fe~{\sc x}, Fe~{\sc xi} up to  Fe~{\sc xxiii}, O~{\sc v} and {\sc vi}, and 
He~{\sc ii}.  We will discuss only the strongest ones, i.e.  He~{\sc ii}~256.32~\AA\ (4.7~K) 
and Fe~{\sc xii}~195.12~\AA\  (6.1~K).  The XRT \citep{Golub2007} was observing with the Al poly 
filter in a field-of-view (FOV) of 384\arcsec $\times$ 384\arcsec\ with a cadence of 30 sec. 
The SUMER spectrometer \citep{Wilhelm1995} took a large raster followed by sit-and-stare observations 
using the 1\arcsec$\times$300\arcsec\ slit in O~{\sc v}~629.77~\AA, N~{\sc v}~1238.82~\AA, 
Mg~{\sc x}~624.90~\AA\ and several other chromospheric lines (C~{\sc i}, Si~{\sc ii} and S~{\sc ii}). 
The O~{\sc v} 629~\AA~line was used to obtain relative Doppler shift maps with the rest wavelength 
taken  as an average from the whole dataset. 

 \begin{figure}[ht!]
 \centering
 \vspace{14.2cm}
\includegraphics{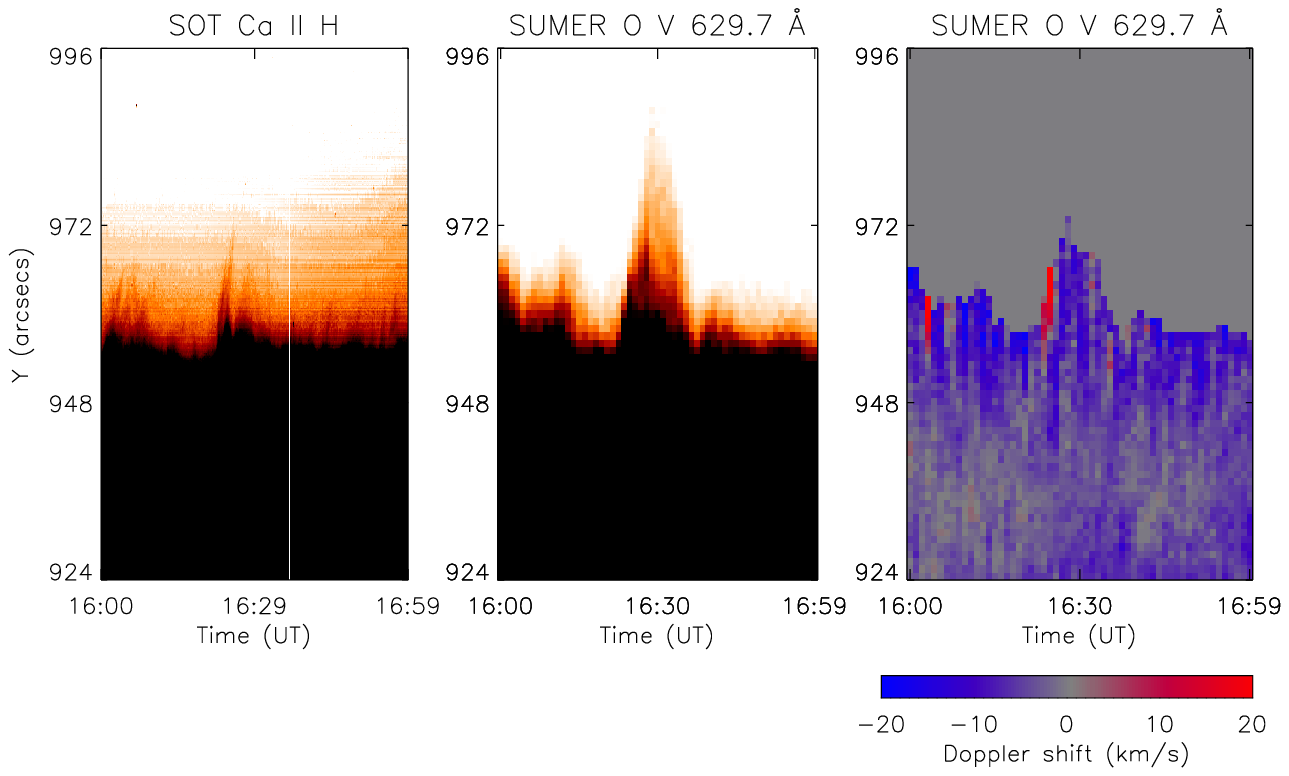}
\includegraphics{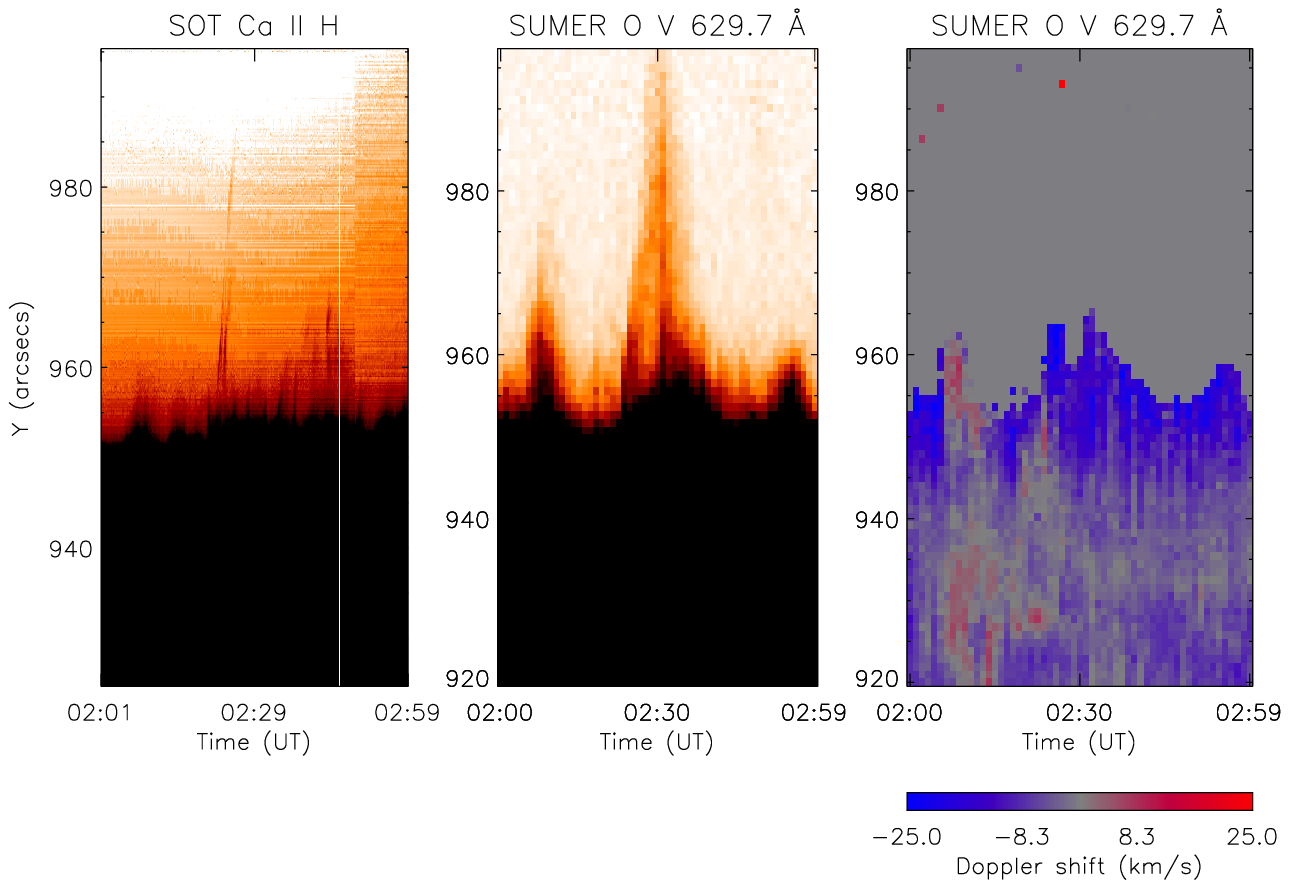}
\includegraphics{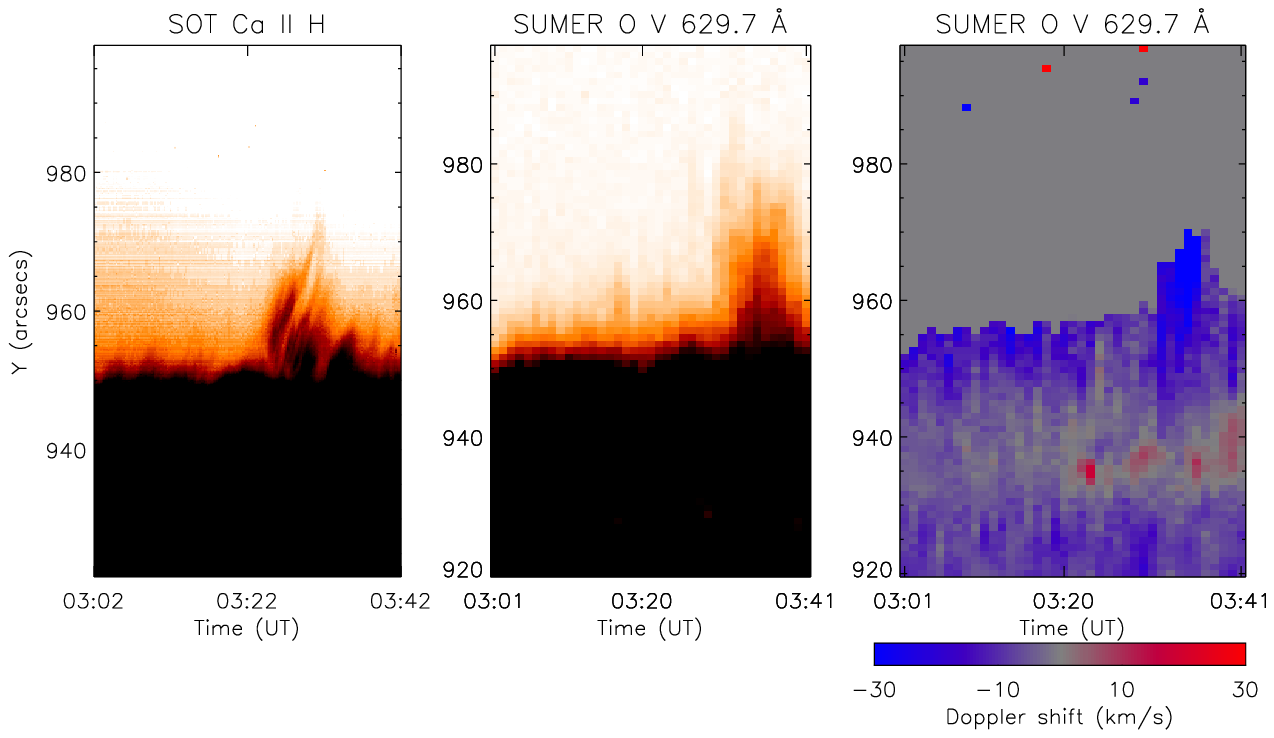}
\vspace{-0.5cm}
 \caption{{\bf Left to right:} SOT Ca~{\sc ii}~H slice-time images, the corresponding SUMER sit-and-stare radiance and Doppler velocity O~{\sc v}  images on April 28 (top) and April 29 (middle and bottom).}
 \label{fig2}
 \end{figure}


Aligning data from various instruments was based on images obtained in spectral lines or filters 
with similar temperatures. First, a SUMER Mg~{\sc x} sit-and-stare image was aligned with a 
co-temporal slice-time image from XRT. Having done this, SUMER and SOT images were then aligned. 
Approximately nine pixels from an SOT image correspond to 1\arcsec\ which is equivalent to the 
SUMER slit width. An automated procedure was written to cut a slice of 9 pixels in time from the SOT 
images and compare them with a SUMER sit-and-stare image in the Si~{\sc ii}, N~{\sc v} and 
O~{\sc v} lines. Due to the uncertainties of the instrument pointing, it is necessary to search 
for the real position using an observed region of at least $\pm$20\arcsec\ from the commanded 
coordinates. We were satisfied with our alignment only when all off-limb features seen 
in the SUMER observations were identified with their SOT Ca~{\sc ii}~H counterpart. The EIS 
data were easily  aligned with XRT using EIS images taken in high-temperature lines. The alignment was done (Fig.~\ref{fig1}) with a precision of $\pm$2\arcsec\
to $\pm$4\arcsec\ which for the size of the observed events (more than 25\arcsec\ in Solar X) is in  
reasonable limits.

 \begin{figure}[ht]
 \centering
 \vspace{8.0cm}
\includegraphics{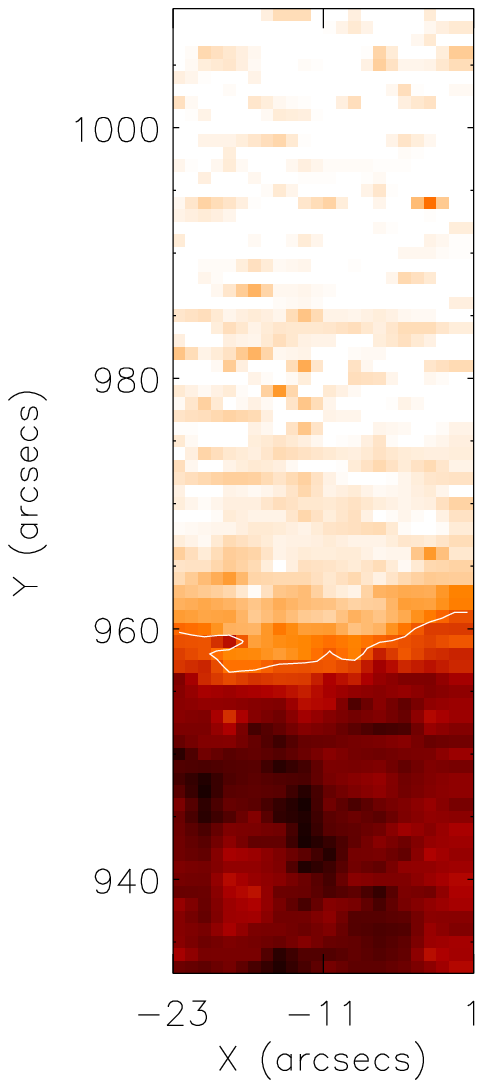}
\includegraphics{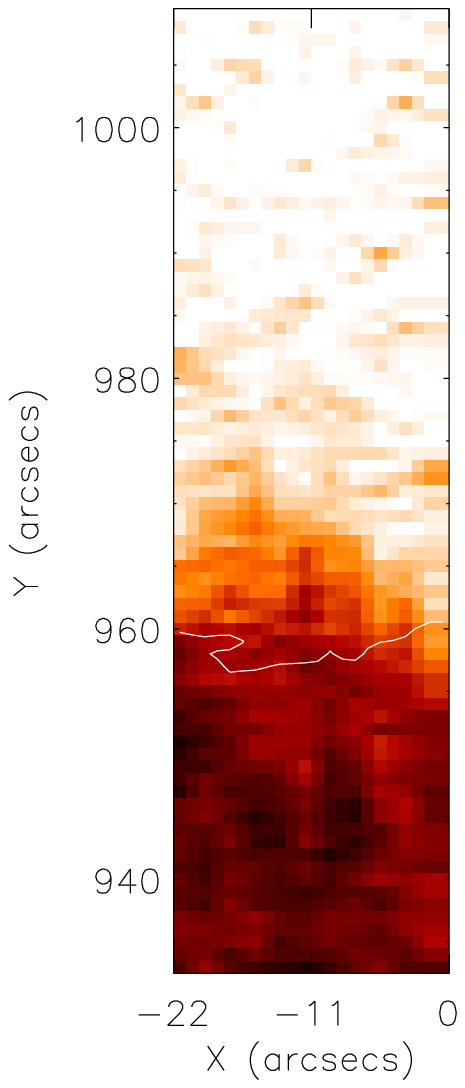}
\includegraphics{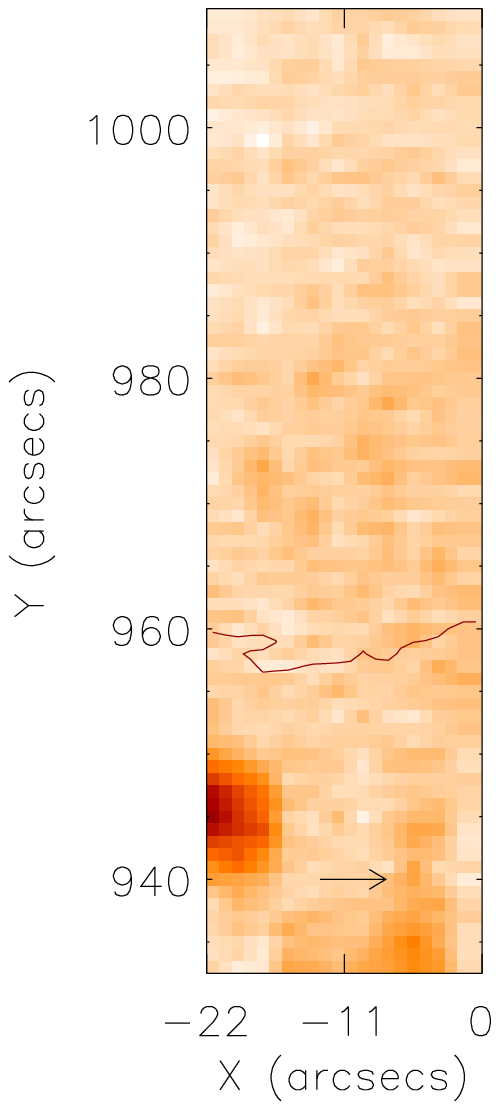}
\includegraphics{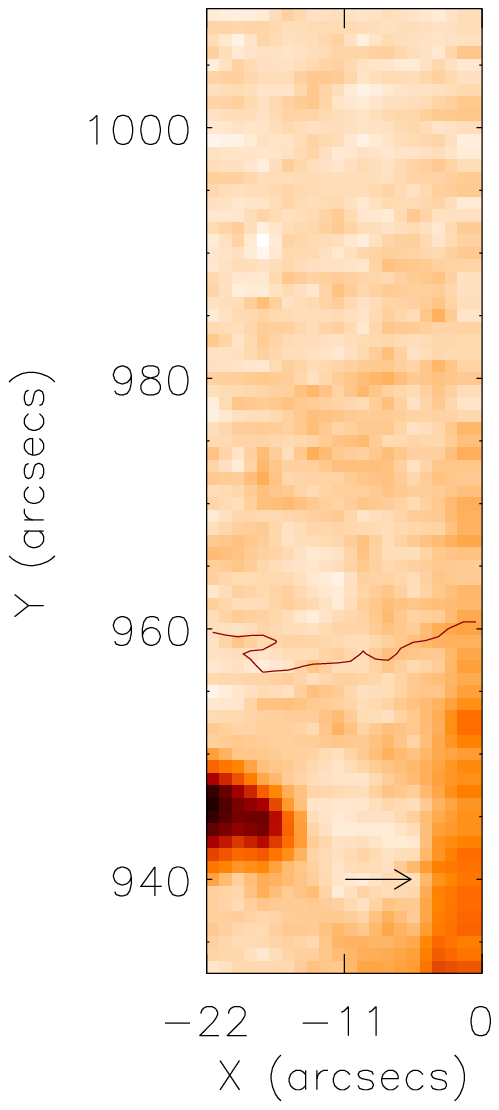}
\vspace*{-5.2cm}
 \caption{EIS rasters taken on April 28 starting at 15:28~UT and 16:23~UT in the 
 He~{\sc ii} 256~\AA\ line (first and second image) and the Fe~{\sc xii} 
 195~\AA\ line (third and fourth image). The arrows indicate the X-ray jet from a coronal bright point 
 identified from the XRT images.}
 \label{fig3}
 \end{figure}

 \begin{figure}[ht!]
 \centering
 \vspace{5.cm}
\includegraphics{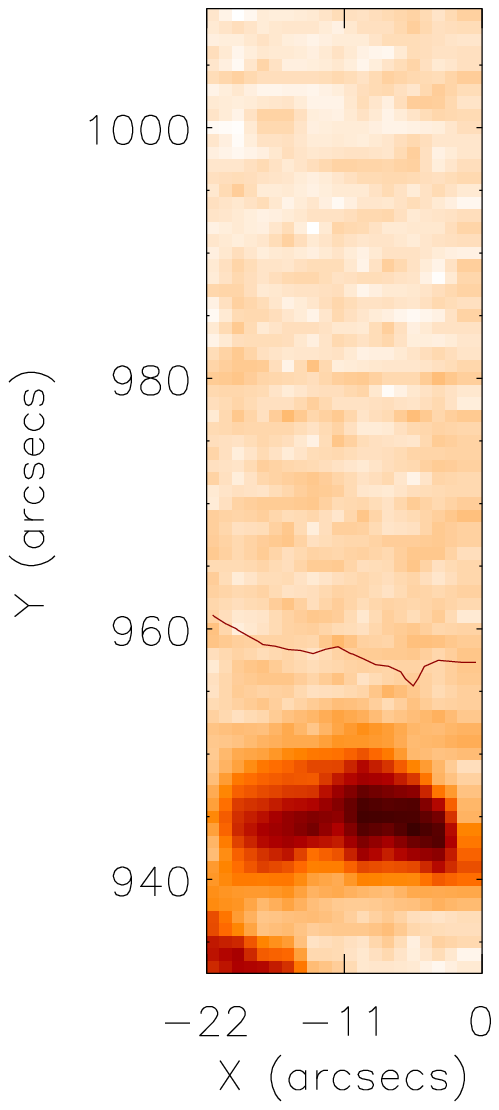}
\includegraphics{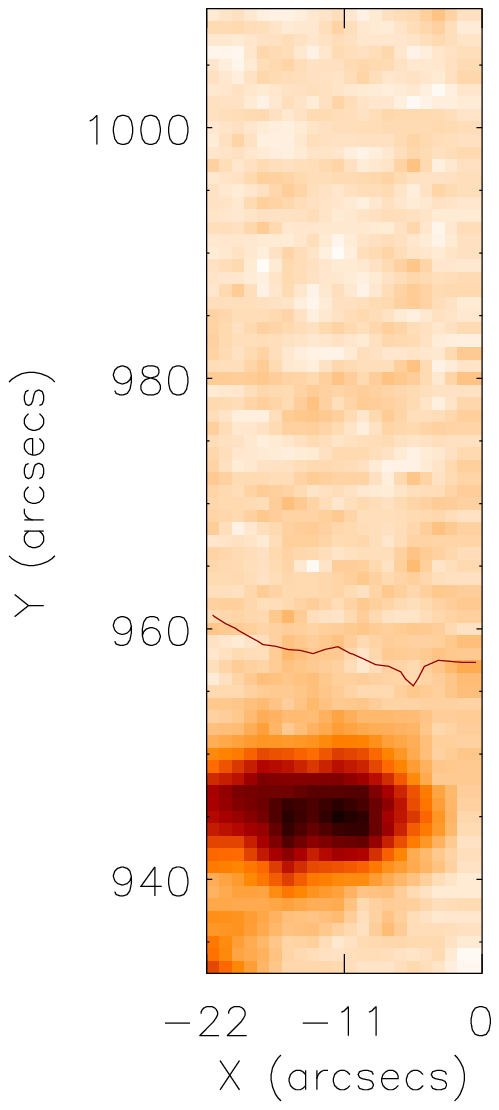}
\includegraphics{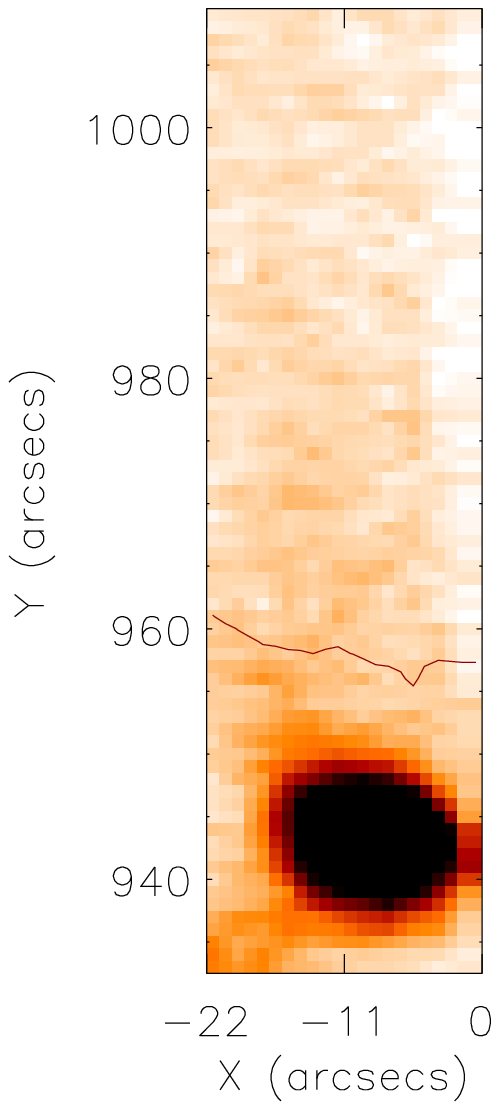}
\vspace{1.5cm}
\includegraphics{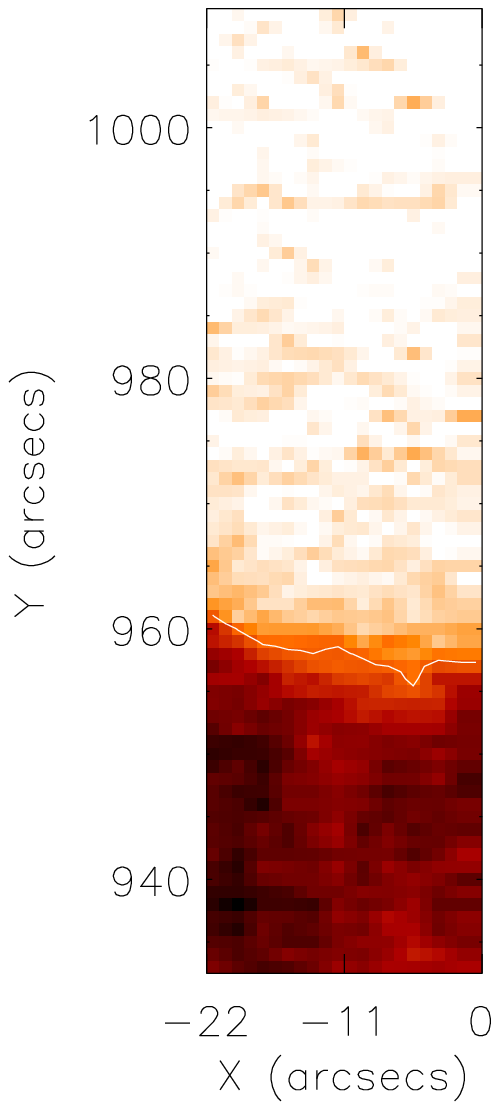}
\includegraphics{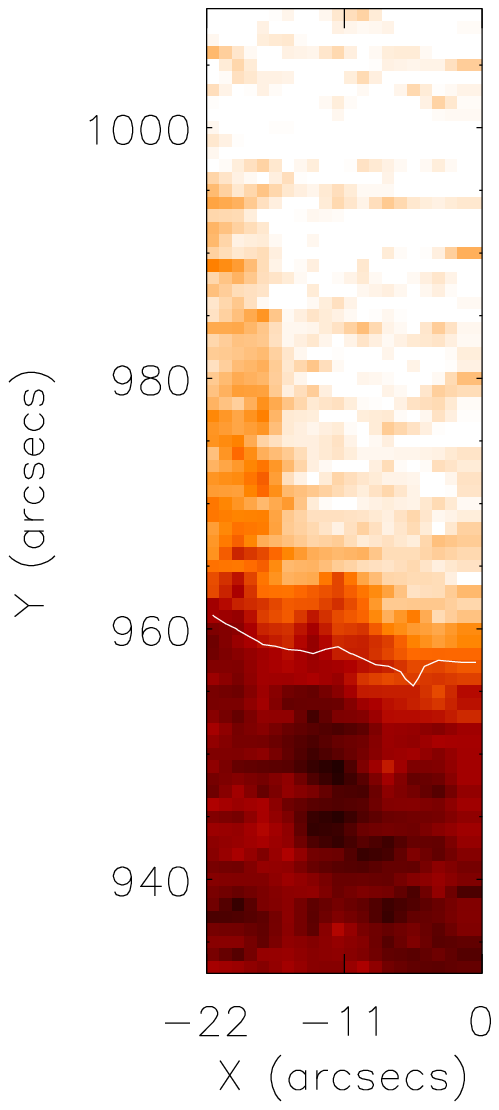}
\includegraphics{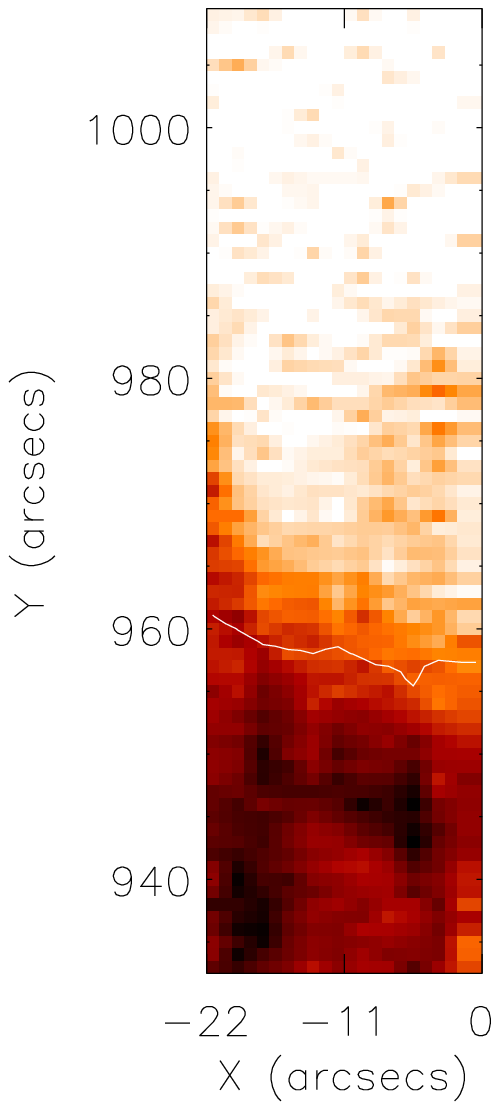}
 \caption{EIS rasters taken on April 29 at 02:09~UT (left), 02:19~UT (middle) and 03:23~UT (right). The top and bottom row images were taken in the He~{\sc ii} 256~\AA\  and  Fe~{\sc xii} 195~\AA\ lines, respectively.}
 \label{fig4}
 \end{figure}

\section{Results}

The observed regions are seeded with thin dynamic jet-like events which 
correspond to the \citet{2007PASJ...59S.655D,2011Sci...331...55D} type II spicules observed in coronal holes (see the online movies). After a very challenging but successful alignment of all instrument data, we selected for further 
analysis the three largest events which fall under the SUMER slit. They are comprised of numerous high velocity thin long spicules (termed 
type II spicules) which evolve simultaneously.  The reason for analysing 
only the largest spicules is the lower spatial  resolution of SUMER, EIS and XRT with respect to SOT 
(for more details see Sect.~2). The ratio of the spatial resolution of  EIS (2\arcsec\ slit width) to SOT (0.1089\arcsec), is approximately 18, which means that  a feature identified in 
SOT has to be larger by more than twice the spatial resolution of EIS, i.e. more than 4\arcsec\ in 
width, in order to be identified in EIS data. In Fig.~\ref{fig1}  we show the three spicules 
which represent  `bushes' of many thin spicules which rise simultaneously forming a large spicule 
(see the online movies). They reach various heights (in average more than 20\,000 km) above the  
solar limb. Their bush-like horizontal expansion reaches up to 25\,000~km. The proper motion of 
the spicules was estimated by following individual spicules in as many images as possible. We 
found a speed of rising between 50 and 250 \kms. The descending velocity was difficult to estimate 
because of the superposition of rotation, up and down-flows. We found a persistent blue-shifted 
emission of less than 50~\kms\ 
even when rotation is not present which apparently corresponds to a down-flow.  
  
The event on April 28  happened at around 16:25~UT and had a duration of 10~min in SOT Ca~{\sc ii}.  
It was registered in SUMER O~{\sc v} 2 min later, i.e. 16:27~UT, and lasted  around 12~min. The 
phenomenon can be described as multiple thin jets which are seen to shoot up at high speed. 
They rotate as one unit as they rise and finally disappear into the surroundings. The first 
event of April 29 took place at around 02:25~UT. The jet-like event rose almost from a blast 
and expended beyond the SOT FOV, i.e. more than 40\,000 km above the solar limb. Just like the 
previous event, it also rotated and then disappeared. The third spicule rose quickly at around 
03:21~UT and sustained its height, rotating for a period longer than the previously 
discussed events before it descended. Again, we observed a delay in the appearance of the event 
in the SUMER O~{\sc v} line. The duration of both events on April 29 is approximately 10 min. 

We analysed the SUMER data taken  in the chromospheric Si~{\sc ii}~1250.41~\AA\ blended with 
the C~{\sc i}~1250.42 and S~{\sc ii}~1250.58~\AA\ lines, the transition region 
O~{\sc v}~629.76~\AA\ and N~{\sc v}~1238.82~\AA\ lines, and the coronal Mg~{\sc x}~624.9~\AA\ line. 
The spicules are clearly present in the chromospheric lines. In these lines as well as in  
transition region lines, the spicules appear as the so-called `macrospicules', i.e. larger size with 
respect to the Ca~{\sc ii}~H spicules.  In addition to the largest spicules discussed here, one can see several small size SUMER spicules which are also composed of several thin long SOT Ca~{\sc ii} spicules. The features can be seen evolving under the SUMER slit  in the edge enhanced image movies). In the transition region lines the spicules show strong 
blue- and red-shifted emission (Fig.~\ref{fig2} and the online movies). For the first time we were able to establish  
the meaning of the blue and red-shifted emission at this line-of-sight position. From 
the comparison of the simultaneously taken SUMER and SOT data (Fig.~\ref{fig2}), we found that 
the Doppler shifts which were earlier interpreted by \citet{1984AdSpR...4...59C, 1997SoPh..175..457P} and \citet{2002A&A...384..303P}
 as rotating motions,  do indeed mean a rotation, but they also 
mean an up-flow which corresponds to a red-shifted emission and a blue-shift which is 
consistent with down-flow. The blue-shift is persistent once the event took off. We should note 
that a single Gauss fit as presented in Fig.~\ref{fig2} does not really give a full picture of 
the dynamics of these events. Only a careful analysis of individual pixels with sufficient 
signal provides the true evolution of the spicule plasmas.

The next step of our study was the analysis of spectral lines formed at coronal temperatures. 
No signal was detected in the coronal SUMER Mg~{\sc x} line for the duration of the large spicules. 
However, due to the uncertainties of using this line for coronal diagnostics 
\citep{2011A&A...526A..19M}, we omit this observation from our conclusions. We concentrated our 
analysis on the behaviour of the EIS spectral lines which are perfectly suitable for coronal 
diagnostics. In order to identify the spicules in EIS data, we used the He~{\sc ii}~256.32~\AA\ 
line as reference. In Fig.~\ref{fig3} we present  a raster taken at 15:28~UT (the last 
good raster before the large spicules took off) on April 28 with no visible  off-limb features. 
By taking a limb contour in the He~{\sc ii}~256~\AA, we established our reference line which 
will represent the solar limb. The presence of the spicule in the  He~{\sc ii} raster image 
(the EIS FOV is shown in Fig.~\ref{fig1}), is more than evident. The large spicule is also seen 
in the O~{\sc v} line, but the low signal in this line does not permit to show a representative 
image. The analysis of all  iron lines from Fe~{\sc viii} up to Fe~{\sc xvi} shows no evidence 
for the presence of a spicule at their corresponding formation temperatures. In Fig.~\ref{fig3} 
(third and fourth image), we give example raster images in the strongest line in the EIS spectrum, the 
Fe~{\sc xii}~195.12~\AA\ line, before (third) and during (fourth) the course of the SUMER spicule.  
The jet-like features seen  in the Fe~{\sc xii} raster images correspond to a X-ray jet from a 
bright point which is situated below the limb. This has been established by using an XRT animated 
sequence. The same is shown in Fig.~\ref{fig4} for the two events on April 29. Again, no trace 
of the large spicule is seen in the coronal lines. 

 \section{Discussion and Conclusions}

The present paper brings out the analysis of three large spicules named as macrospicules as seen 
in EUV spectral lines. The macrospicules appear to consist of many thread-like spicules as seen 
in high-resolution SOT Ca~{\sc ii} H images which rise, rotate, descend and in general evolve  
almost simultaneously. We identified  the counterpart of these large spicules in SUMER and EIS 
sit-and-stare and raster images, respectively. The large SOT spicules are very dynamic and  
appear as features which are usually identified as macrospicules in transition region emission 
lines. Our main result is that these spicules although very large and  dynamic, show no 
presence in spectral lines formed at temperatures above 300\,000~K.   Macrospicules have been 
reported in coronal lines  \citep{1997SoPh..175..457P} and recently, a feature called macrospicule 
identified in the EIS He~{\sc ii} line was clearly seen as an X-ray jet  \citep{2010A&A...510L...1K}. 
Where does the discrepancy between our results and these observations  comes from? We strongly believe 
that the mis-identification or rather the identification of a different type of phenomena as  
macrospicules is due to the interpretation of the transition region emission and especially, the  He~{\sc i} and {\sc ii} emission. 
Our recent multi-instrument  analysis of a `classical' X-ray jet clearly 
demonstrated that X-ray jets are strongly visible in the He~{\sc ii} emission 
\citep{2011A&A...526A..19M}. We believe that this is also the case of the event described by   
\citet{2010A&A...510L...1K}. Equally, spicules are prominent in these spectral lines.  
Forthcoming papers by Madjarska et al. (2011, in prep) and Subramanian et al. (2011, in prep) will give more 
information on this subject. 

 Recently, \citet{2011Sci...331...55D} reported the observation of spicules in the AIA/SDO~171~\AA\ channel in active regions,
 quiet Sun as well as  in coronal holes (Fig. 3 in their paper and movies S8 and S9).
This fact made us ask: Why do we not detect any coronal emission in EIS data of coronal hole spicules? It should 
be noted here that the two works (ours and the work of De Pontieu and co-authors) deal with different type of observations, spectroscopic and 
imaging. So, we decided to make a preliminary spectroscopic analysis of the AIA ~171~\AA\ channel. 
This channel will be dominated 
by Fe~{\sc ix} emission (http://aia.lmsal.com/public/results.htm). However, it also contains an abundance of cool Fe~{\sc viii}, Ne~{\sc v}, 
O~{\sc v} and {\sc vi} lines and therefore, when along the line-of-sight there is no emission 
from the Fe~{\sc ix} lines, the cooler lines will have a significant contribution. This was 
clearly demonstrated from simultaneous TRACE~Fe~{\sc ix/x}~171~\AA, and SUMER transition region 
and coronal observations \citep[][see Sect. 4]{2008A&A...482..273M}. In order to make a 
preliminary analysis of the cooler emission contribution in the AIA~171, we checked co-observations 
of a large prominence by EIS and AIA. Solar prominences have very similar plasma 
parameters as spicules concerning temperatures and densities, but with the prominences being much 
larger, especially  quiescent ones. Surprisingly, we identified clearly all solar 
prominences in the AIA 171~\AA\ images. A check on the EIS observations 
for one of these prominences revealed an emission not higher than Fe~{\sc viii} with a  
contribution function peaking  at 400\,000~K, i.e. at transition region temperature. Is it, 
therefore, possible that the spicules seen by  \citet{2011Sci...331...55D} in the AIA~171~\AA\ 
channel come from the emission of lower temperature lines? This question needs to be answered 
by careful spectroscopic studies  of the response of the AIA channels.

\begin{acknowledgements} The authors thank ISSI for the support of the 
team ``Small-scale transient phenomena and their contribution to coronal 
heating''. Research at Armagh Observatory is grant-aided by the N.~Ireland 
Department of Culture, Arts and Leisure. We also thank STFC for support via 
grants ST/F001843/1 and ST/H001921/1. Hinode is a Japanese mission developed 
and launched by ISAS/JAXA, with NAOJ as domestic partner and NASA and STFC (UK) as international partners.
 \end{acknowledgements} 
    
\bibliography{refe_intro}

\end{document}